\documentclass{article} % For LaTeX2e
\usepackage[preprint]{colm2026_conference}

\usepackage{microtype}
\usepackage{hyperref}
\usepackage{url}
\usepackage{booktabs}
\usepackage{graphicx}
\usepackage{amsmath,amssymb}
\usepackage{xcolor}
\usepackage{xspace}
\usepackage{algorithm}
\usepackage{algorithmic}
\usepackage{multirow}
\usepackage{subcaption}
\usepackage{enumitem}
\usepackage{lineno}

\definecolor{darkblue}{rgb}{0, 0, 0.5}
\hypersetup{colorlinks=true, citecolor=darkblue, linkcolor=darkblue, urlcolor=darkblue}

\graphicspath{{figs/}}

\definecolor{weikaiGreen}{HTML}{196F3D}

\ifx\figurename\undefined \def\figurename{Figure}\fi
\renewcommand{\figurename}{Fig.}

\newcommand{\Sect}[1]{Section~\ref{#1}}
\newcommand{\Fig}[1]{Fig.~\ref{#1}}
\newcommand{\Tbl}[1]{Table~\ref{#1}}

\newcommand{\Apx}[1]{Appendix~\ref{#1}}

\newcommand{\proj}{\textsc{SafeDream}\xspace}
\newcommand{\no}[1]{}

\newcommand{\bz}{\mathbf{z}}
\newcommand{\bh}{\mathbf{h}}
\newcommand{\bs}{\mathbf{s}}
\newcommand{\ba}{\mathbf{a}}
\newcommand{\Gt}{G_t}
\newcommand{\Vt}{V_t}

\title{SafeDream: Safety World Model for Proactive Early Jailbreak Detection}

\author{Bo Yan \\
University of Central Florida \\
\texttt{bo949643@ucf.edu}
\And
Weikai Lin \\
University of Rochester \\
\texttt{wlin33@ur.rochester.edu}
\And
Yada Zhu \\
IBM Research\\
\texttt{yzhu@us.ibm.com}
\And
Song Wang\thanks{Corresponding author.} \\
University of Central Florida \\
\texttt{song.wang@ucf.edu}}

\begin{document}

\ifcolmsubmission
\linenumbers
\fi

\maketitle

\begin{abstract}
Multi-turn jailbreak attacks progressively erode LLM safety alignment across seemingly innocuous conversation turns, achieving success rates exceeding 90\% against state-of-the-art models.
Existing alignment-based and guardrail methods suffer from three key limitations: they require costly weight modification, evaluate each turn independently without modeling cumulative safety erosion, and detect attacks only \emph{after} harmful content has been generated.
To address these limitations, we first formulate the \emph{proactive early jailbreak detection} problem with a new metric, \emph{detection lead}, that measures how early an attack can be detected before the LLM complies.
We then propose \proj, a lightweight world-model-based framework that operates as an external module without modifying the LLM's weights.
\proj introduces three components: (1)~a \emph{safety state world model} that encodes LLM hidden states into a compact safety representation and predicts how it evolves across turns, (2)~\emph{CUSUM detection} that accumulates weak per-turn risk signals into reliable evidence, and (3)~\emph{contrastive imagination} that simultaneously rolls out attack and benign futures in latent space to issue early alarms before jailbreaks occur.
On three multi-turn jailbreak benchmarks (XGuard-Train, SafeDialBench, SafeMTData) against 8 baselines, \proj achieves the best detection timeliness across all benchmarks (1.06--1.20 turns before compliance) while maintaining competitive false positive rates and outperforming baselines in detection quality.

\end{abstract}

\begin{figure*}[h!]
    \centering
    \includegraphics[width=\linewidth]{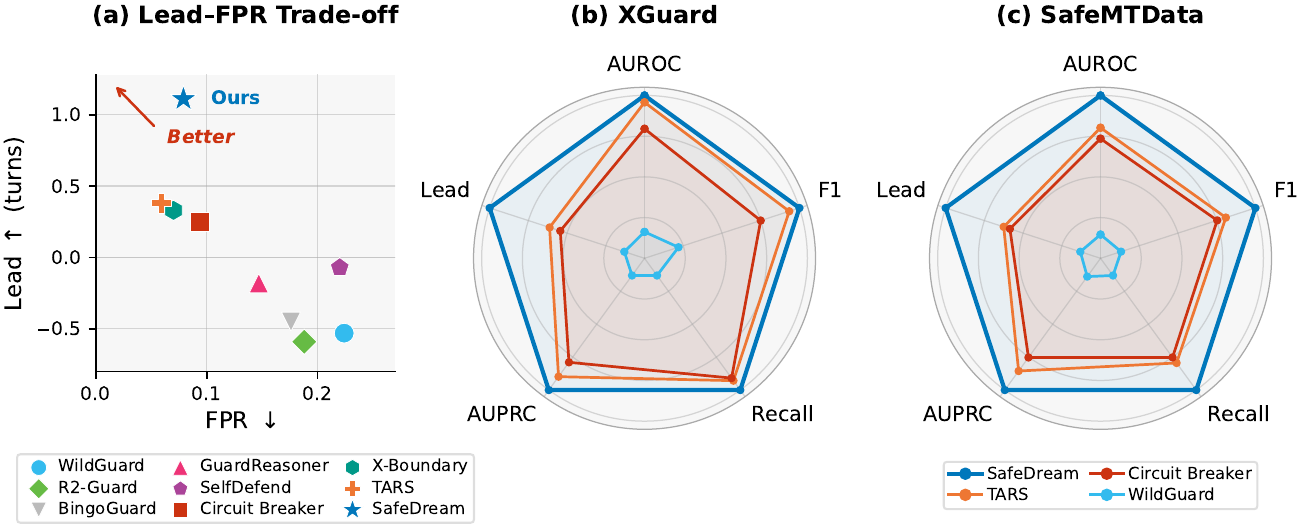}
    \caption{\textbf{\proj achieves the best Lead--FPR trade-off and outperforms all baselines across benchmarks.} (a)~Lead vs.\ FPR on XGuard: Lead measures how many turns earlier a jailbreak is detected, and FPR is the false alarm rate. \proj (blue star) achieves the best trade-off between Lead and FPR. (b,c)~Radar charts on XGuard and SafeMTData comparing \proj against baselines across five metrics (axes individually normalized for readability). \proj dominates all baselines on all metrics across both benchmarks.}
    
    \label{fig:teaser}
\end{figure*}

\section{Introduction}
\label{sec:intro}
Large language models (LLMs)~\citep{qwen2.5} can be exploited for harmful purposes such as generating misinformation or illegal instructions.
Despite safety mechanisms like alignment training and guardrails~\citep{llamaguard, wildguard2024}, multi-turn jailbreak attacks remain highly effective~\citep{harmbench2024}: by spreading malicious intent across seemingly innocuous turns, they gradually erode LLM safety barriers until the model complies.
Crescendo~\citep{russinovich2025crescendo} achieves a 98\% binary success rate against GPT-4, FITD~\citep{fitd2025} averages 94\% Attack Success Rate (ASR) across seven LLMs, and STAR~\citep{star2026} reaches 94.5\% Safety Failure Rate (SFR) against GPT-4o.

Recent work has explored various approaches to counter jailbreak attempts.
Alignment-based methods such as Circuit Breaker~\citep{circuitbreaker2024}, X-Boundary~\citep{xboundary2025}, and TARS~\citep{tars2025} modify LLM weights to resist attacks.
Guardrail methods such as WildGuard~\citep{wildguard2024} and GuardReasoner~\citep{guardreasoner2025} deploy external classifiers to evaluate each turn independently.
However, these approaches suffer from three key limitations.
First, alignment-based methods require access to and modification of model weights, incurring significant computational overhead and precluding deployment on closed-source models.
Second, both paradigms typically evaluate each turn independently, making them unable to defend against multi-turn attacks where the LLM complies only after a \emph{gradual safety erosion} accumulated across multiple turns.
Third, neither paradigm formulates the problem of \emph{anticipating} jailbreaks before they succeed; by the time these methods detect an attack signal, the LLM may have already generated harmful content.

To address these three limitations, we reformulate the multi-turn jailbreak detection problem (\Sect{sec:formulation}).
Unlike previous formulations, we explicitly incorporate \emph{detection lead time}, measuring how many turns earlier a jailbreak attempt can be detected before the LLM actually complies.
We then propose \proj, a world-model-based safety module that enables proactive, early jailbreak detection in multi-turn dialogues.
\proj addresses the three limitations above:
it operates as a lightweight external module without modifying the LLM's weights (\emph{solving limitation 1});
it models safety state dynamics across turns rather than evaluating each turn independently (\emph{solving limitation 2});
and it proactively detects attacks before the LLM complies through contrastive imagination (\emph{solving limitation 3}).

\proj consists of three key components.
First, we introduce a \emph{Safety State World Model}: a lightweight Transformer that predicts future safety states from the safety state history and user actions, operating entirely in a low-dimensional safety state space (\Sect{sec:world_model}).
Second, we employ \emph{Cumulative Sum (CUSUM) detection} to aggregate risk signals across multiple turns, demonstrating how multi-turn safety erosion can be effectively captured through sequential evidence accumulation (\Sect{sec:cusum}).
Third, in ambiguous situations, in order to detect jailbreaks as early as possible, we use our Safety State World Model to conduct \emph{contrastive imagination}, which rolls out possible attack and benign futures and enables early alarms before a jailbreak actually occurs (\Sect{sec:imagination}).

We evaluate \proj on three multi-turn jailbreak benchmarks (XGuard-Train~\citep{xteaming2025}, SafeDialBench~\citep{safedialbench2025}, SafeMTData~\citep{actorattack2025}) against 8 baselines.
Our key contributions are:
\begin{itemize}[leftmargin=0.2in]
    \item We formulate the \emph{proactive early jailbreak detection} problem with a new metric, \emph{detection lead}, that measures how early attacks can be detected before the LLM complies.
    \item We propose \proj, a world-model-based framework consisting of a safety state world model, CUSUM detection, and contrastive imagination for proactive early jailbreak detection.
    \item \proj achieves the best detection lead across all benchmarks (1.06--1.20 turns before compliance) and the highest detection quality with only an external module, even outperforming baselines that modify LLM weights.
    \item Ablation studies confirm the effectiveness of our design: contrastive imagination is the primary driver of early detection and CUSUM is the key for false alarm suppression.
\end{itemize}

\begin{figure*}[t]
    \centering
    \includegraphics[width=\linewidth]{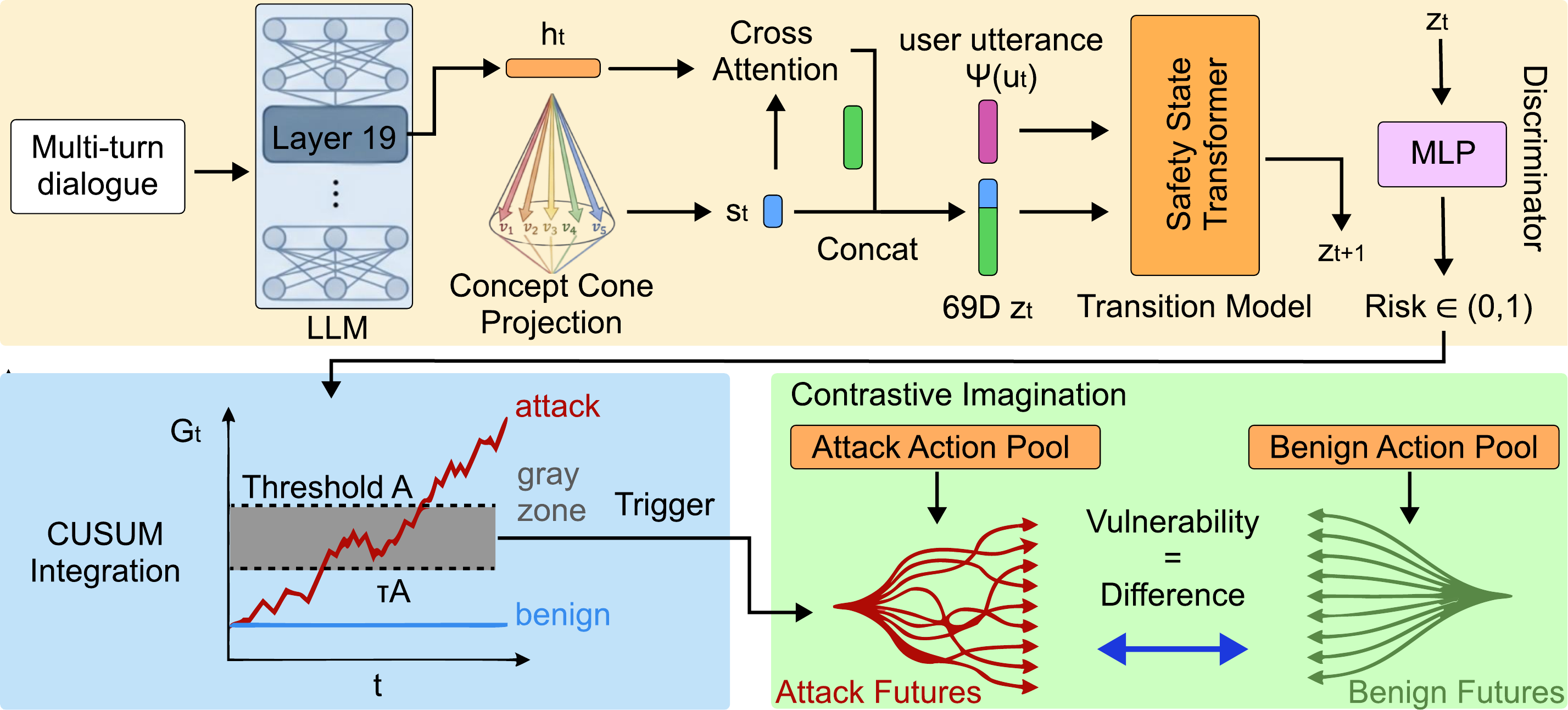}
    \caption{\textbf{\proj overview.}
    At each turn $t$, the frozen LLM provides hidden-state observations $h_t$, which are projected into a safety-grounded latent space $s_t$ via the concept cone.
    The safety state $z_t$ is obtained by enriching $s_t$ through cross-attention.
    A lightweight Transformer learns transition dynamics by predicting $z_{t+1}$ conditioned on user actions.
    A discriminator maps each $z_t$ to a risk score $r_t$.
    CUSUM accumulates risk scores across turns to detect safety state shifts; when evidence is ambiguous (gray zone), contrastive imagination rolls out both attack and benign futures to quantify vulnerability and enable early detection.}
    \label{fig:overview}
\end{figure*}

\section{Problem Formulation}
\label{sec:formulation}
\paragraph{Multi-turn dialogue and safety model.}
We study jailbreak detection given an ongoing dialogue between a user and an LLM, the goal is to decide whether the user is steering the conversation toward harmful content \emph{before} the LLM complies.
At turn $t$, the user sends utterance $u_t$ and the LLM generates response $y_t$.
We define the observation $o_t = (u_1, y_1, \ldots, y_{t-1}, u_t)$ as the conversation prefix immediately before $y_t$ is generated.
A safety model $f(o_t) \in \{0, 1\}$ infers whether jailbreak intent is present; if $f(o_t) = 1$, the LLM refuses to respond.

\paragraph{Attack model and compliance point.}
In a multi-turn jailbreak, an adversarial user progressively guides the LLM toward harmful content through seemingly innocuous messages, gradually eroding the LLM's safety alignment until the model fully complies.
We call this critical point the \emph{compliance point} $c$, formally defined as the first turn at which the majority of independent safety classifiers flag the LLM's response as harmful (see \Sect{sec:training} for the labeling procedure).
Detecting the attack after the compliance point ($t > c$) is too late, as harmful content has been generated.

\paragraph{Proactive early jailbreak detection.}
Previous defenses show little concern for \emph{how early} an attack can be detected: guardrails~\citep{wildguard2024, guardreasoner2025} flag harmful content only after it appears, and alignment-based methods~\citep{circuitbreaker2024, tars2025} resist attacks but do not anticipate compliance.
We formulate the \emph{proactive early jailbreak detection} problem: the safety model $f$ predicts whether the dialogue will eventually reach a compliance point, and we measure timeliness via the \emph{detection lead}:
\begin{equation}
    \text{Lead} = c - t^{\text{det}}, \quad \text{where } t^{\text{det}} = \min\{t : f(o_t) = 1\}.
\end{equation}
Here, $t^{\text{det}}$ is the earliest turn when the model detects a risk.
Positive Lead ($t^{\text{det}} < c$) achieves proactive early detection; negative Lead indicates reactive post-hoc detection.
The objective is to maximize Lead while maintaining a low false positive rate on benign conversations.

\section{SafeDream: Proactive World Model  for Early Jailbreak Detection}
\label{sec:method}

We present \proj, a world-model-based framework for proactive early jailbreak detection (\Fig{fig:overview}).
The framework introduces a \emph{safety state world model} (\Sect{sec:world_model}) that encodes the LLM's hidden state into compact latent safety states and learns to predict future safety states.
To reliably detect jailbreak, we employ \emph{cumulative sum detection} (\Sect{sec:cusum}), which aggregates weak per-turn risk signals into robust statistical evidence.
Building on the world model and cumulative detection, we propose \emph{contrastive imagination} (\Sect{sec:imagination}), which simultaneously predicts attack and benign future roll-outs to quantify vulnerability, enabling proactive early attack detection.
We describe the training procedure in \Sect{sec:training}.

% ======================================================
% 4.1 Safety State World Model
% ======================================================
\subsection{Safety State World Model}
\label{sec:world_model}

Proactive early detection requires the ability to predict future safety outcomes from the current state.
To this end, we design a world model over safety states, consisting of two components:
(1)~a \emph{safety state extractor} $e_\phi$ that compresses each turn into a low-dimensional safety state vector $\bz_t = e_\phi(o_t)$, capturing the safety-relevant information of that turn;
and (2)~a \emph{transition model} $g_\theta$ that integrates safety states from previous turns and predicts future safety states based on user actions $\ba_t = \psi(u_t)$.
Each safety state is converted to risk score by a discriminator $d_\omega(\bz_t) \to r_t \in [0, 1]$, where a value of 1 indicates likely jailbreak compliance.

\paragraph{Safety state extraction.}
At each turn $t$, we feed the conversation prefix $o_t$ through the frozen Qwen2.5-7B~\citep{qwen2.5} and extract the hidden state $\bh_t \in \mathbb{R}^{3584}$ from layer 19, which \citet{conceptcone2025} identify as the layer most strongly associated with safety-related refusal behavior in Qwen-family models.
To extract safety-relevant information, we project $\bh_t$ onto $K{=}5$ safety-related directions on the concept cone~\citep{conceptcone2025}, yielding a safety signature $\bs_t \in \mathbb{R}^5$.
We further enrich $\bs_t$ via Perceiver-style cross-attention~\citep{perceiverio2022} that uses $\bs_t$ as queries into $\bh_t$, producing a learned 64-dimensional extension.
The final safety state is:
\begin{equation}
    \bz_t = [\bs_t \;;\; \text{XAttn}_\phi(\bs_t, \bh_t)] \in \mathbb{R}^{69},
\end{equation}
where the first 5 dimensions are frozen cone projections and the remaining 64 are learned enhancement through $\phi$.
Architectural details are in \Apx{sec:app_architecture}.

\paragraph{Transition model.}
The user action $\ba_t = \psi(u_t)$ is encoded by embedding the utterance $u_t$ with the frozen LLM, mean-pooling over token embeddings, and linearly projecting the result to a 64-dimensional vector.
A causal Transformer $g_\theta$ then models the safety-state dynamics~\citep{dreamerv3}.
The goal of the transition model is to predict the next safety state from the current state and user action: given the input token $[\bz_{t-1}; \ba_t]$, the model outputs the predicted $\hat{\bz}_t$:
\begin{equation}
    \hat{\bz}_{t} = g_\theta\big((\bz_0, \ba_1), (\bz_1, \ba_2), \ldots, (\bz_{t-1}, \ba_t)\big),
    \label{eq:transition}
\end{equation}
where $\bz_0 = \mathbf{0}$ is a fixed initial state.
The Transformer has only 1.2M parameters (0.017\% of the 7B LLM), enabling the imagination module to roll out futures with negligible overhead.

\paragraph{Safety discriminator.}
A two-layer MLP $d_\omega(\bz_t) \to r_t \in [0, 1]$ maps each safety state to a risk score, which is used by the cumulative detection module (\Sect{sec:cusum}) and the contrastive imagination module (\Sect{sec:imagination}).

% ======================================================
% 4.2 Cumulative Sum Attack Detection
% ======================================================
\subsection{Cumulative Sum Attack Detection}
\label{sec:cusum}

In a multi-turn attack, the risk score from any single turn is weak and noisy.
However, if we accumulate risk signals across turns, genuine attacks reveal a \emph{systematic risk increasing trend} that distinguishes them from fluctuations in benign conversations.
We adopt CUSUM~\citep{page1954cusum}, a sequential change-point detection method that is provably optimal for minimizing detection delay at a given false alarm rate~\citep{moustakides1986cusum}.

We first convert each risk score to log-odds via a clipped logit transformation:
\begin{equation}
    \Lambda_t = \mathrm{logit}_\epsilon(r_t) \;\triangleq\; \log \frac{\mathrm{clip}(r_t,\, \epsilon,\, 1{-}\epsilon)}{1 - \mathrm{clip}(r_t,\, \epsilon,\, 1{-}\epsilon)},
    \label{eq:logit}
\end{equation}
with $\epsilon = 10^{-6}$.
The CUSUM statistic accumulates evidence by subtracting a baseline $\kappa$ (the average log-odds under benign conversations):
\begin{equation}
    \Gt = \max(0, \; G_{t-1} + \Lambda_t - \kappa), \quad G_0 = 0.
    \label{eq:cusum}
\end{equation}
In benign conversations, $\Lambda_t \approx \kappa$ and $\Gt$ stays near zero.
During an attack, $\Lambda_t$ consistently exceeds $\kappa$, causing $\Gt$ to climb steadily.
An alarm is raised when $\Gt \geq A$.

We define a \emph{gray zone} ($\tau A < \Gt < A$, $\tau \in (0,1)$) where attack signals have accumulated but the evidence is not yet conclusive.
In this zone, \proj activates contrastive imagination (\Sect{sec:imagination}) to proactively resolve the ambiguity.

% ======================================================
% 4.3 Contrastive Imagination for Early Attack Detection
% ======================================================
\subsection{Contrastive Imagination for Early Attack Detection}
\label{sec:imagination}

When CUSUM enters the gray zone, instead of waiting for more turns, \proj leverages the world model (\Sect{sec:world_model}) to \emph{imagine} what would happen next.
We first construct two action pools from the training data: an attack pool $\mathcal{A}_{\text{atk}}$ containing encoded attack utterances and a benign pool $\mathcal{A}_{\text{ben}}$ containing encoded benign utterances.
For each condition $\gamma \in \{\text{atk}, \text{ben}\}$, we randomly sample actions from the corresponding pool to simulate different futures.
Specifically, we roll out $M{=}8$ trajectories of $H{=}3$ steps from the current state, autoregressively sampling actions, predicting states via $g_\theta$, scoring via $d_\omega$, and accumulating the imagined CUSUM (full procedure in \Apx{sec:app_imagination_procedure}).
After the rollout, we define the \emph{vulnerability score} $\Vt$, which compares the imagined CUSUM endpoints between attack and benign futures:
\begin{equation}
    \Vt = \frac{1}{M}\sum_{k=1}^{M} \hat{G}_{t+H}^{(k,\text{atk})} - \frac{1}{M}\sum_{k=1}^{M} \hat{G}_{t+H}^{(k,\text{ben})}.
    \label{eq:vulnerability}
\end{equation}
When $\Vt$ exceeds a threshold $A_{\text{imag}}$, \proj triggers an early alarm.
The rationale is: if the conversation is genuinely vulnerable, attack futures cause the imagined CUSUM to spike while benign futures keep it stable, producing a large $\Vt$.
If the conversation is safe, both futures produce similar outcomes (small gap).
Empirically, we find that this contrastive design avoids the false alarms that attack-only imagination would produce (\Sect{sec:ablation}).

The full decision rule combines CUSUM and imagination:
\begin{equation}
    \text{Alarm}(t) = \begin{cases}
        1 & \text{if } \Gt \geq A, \\
        1 & \text{if } \tau A < \Gt < A \;\text{ and }\; \Vt \geq A_{\text{imag}}, \\
        0 & \text{otherwise.}
    \end{cases}
    \label{eq:alarm}
\end{equation}
Imagination is triggered in gray zone, adding sub-100\,ms overhead only when needed.

\begin{figure}[t]
    \centering
    \includegraphics[width=0.9\linewidth]{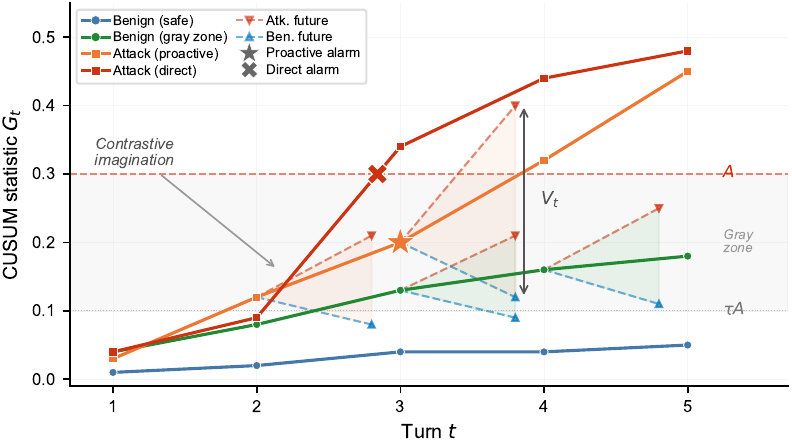}
    \caption{\textbf{CUSUM dynamics and contrastive imagination.} In the gray zone ($\tau A < \Gt < A$), \proj forks the current state into attack futures (red dashed, upward) and benign futures (blue dashed, downward). A large vulnerability gap $\Vt$ triggers a proactive alarm ($\bigstar$) while $\Gt$ is still below $A$. Conversations that jump past the gray zone receive a direct alarm ($\times$). Benign conversations entering the gray zone produce small $\Vt$ and no alarm.}
    \label{fig:cusum}
\end{figure}

\Fig{fig:cusum} illustrates the detection pipeline.
In the gray zone, contrastive imagination forks the current state into attack and benign futures (dashed lines); the gap $\Vt$ between them determines proactive alarm (star marker) \emph{before} $\Gt$ reaches $A$.
Conversations that skip the gray zone trigger a direct alarm (cross marker) when $\Gt \geq A$.
Benign conversations may enter the gray zone, but imagination reveals a small $\Vt$, suppressing false alarms.

% ======================================================
% 4.4 Training
% ======================================================
\subsection{Training}
\label{sec:training}

\paragraph{Labeling.}
We obtain turn-level safety labels through majority voting of three independent classifiers: HarmBench~\citep{harmbench2024}, GPT-4o~\citep{gpt4o2024}, and MD-Judge~\citep{saladbench2024}.
The compliance point $c$ is the earliest turn at which a majority of the three classifiers flag the response as harmful ($\ell_t \geq 0.67$), ensuring that the same threshold governs both training and evaluation.
None of these classifiers appears as an evaluation baseline.

\paragraph{Loss function.}
To enable smooth supervision, we use a soft label $\ell_t = \#\{\text{classifiers flagging harmful}\}/3$.
The total loss combines three terms:
\begin{equation}
    \min_{\theta, \phi, \omega} \;\; \mathcal{L}_{\text{trans}} + \lambda_1 \mathcal{L}_{\text{disc}} + \lambda_2 \mathcal{L}_{\text{imag}},
\end{equation}
where $\mathcal{L}_{\text{disc}} = \sum_{n,t} \mathrm{BCE}(d_\omega(\bz_t^{(n)}), \ell_t^{(n)})$ trains the discriminator on soft labels, $\mathcal{L}_{\text{trans}} = \sum_{n,t} \| \hat{\bz}_t^{(n)} - \mathrm{sg}(\bz_t^{(n)}) \|_2^2$ trains the transition model to predict future safety state, and $\mathcal{L}_{\text{imag}}$ bridges the observed-vs-imagined distribution gap by training $d_\omega$ on imagined trajectories following DreamerV3~\citep{dreamerv3}.
The imagination loss uses a margin term to ensure separation between attack and benign risk scores, plus a calibration term to anchor absolute values (details in \Apx{sec:app_config}).
We set $\lambda_1 = 1.0$, $\lambda_2 = 0.5$.
% Training completes in approximately 30 minutes on a single GPU.

\section{Experiments}
\label{sec:experiments}
\subsection{Experimental Setup}
\label{sec:setup}

\paragraph{Datasets.}
We evaluate on three multi-turn jailbreak benchmarks that cover complementary attack characteristics.
\textbf{XGuard-Train}~\citep{xteaming2025} (30K conversations, 13 risk categories, adaptive multi-agent attacks) serves as our primary benchmark due to its scale and diversity.
\textbf{SafeDialBench}~\citep{safedialbench2025} (4K conversations, 7 distinct attack strategies, bilingual) tests generalization across diverse attack tactics.
\textbf{SafeMTData}~\citep{actorattack2025} (1.7K conversations, actor-network attacks, $\sim$8.7 turns) provides backward compatibility with prior work such as TARS.
For each benchmark, to prevent the model from learning the topic shortcuts for detection, we construct topic-matched benign contrasts to force models to distinguish based on trajectory dynamics rather than topic.

\paragraph{Baselines.}
We compare against 8 baselines spanning two categories (full descriptions in \Apx{sec:app_baselines}):
\begin{itemize}[leftmargin=0.2in]
    \item \textbf{Guardrail methods}: WildGuard~\citep{wildguard2024}, R2-Guard~\citep{r2guard2025}, BingoGuard~\citep{bingoguard2025}, GuardReasoner~\citep{guardreasoner2025}, SelfDefend~\citep{selfdefend2025} --- external classifiers that evaluate each turn independently.
    \item \textbf{Alignment-based methods}: Circuit Breaker~\citep{circuitbreaker2024}, X-Boundary~\citep{xboundary2025}, TARS~\citep{tars2025} --- methods that modify LLM weights to internalize safety.
\end{itemize}

\paragraph{Metrics.}
We report standard detection quality metrics: Area Under the ROC Curve (AUROC), Area Under the Precision-Recall Curve (AUPRC), F1 score, Recall, and False Positive Rate (FPR).
For detection timeliness, we use \textbf{Lead} (detection lead time), the metric we proposed in \Sect{sec:formulation}, which measures the number of turns by which the alarm precedes the compliance point.
Positive Lead indicates proactive early detection; negative Lead indicates reactive post-hoc detection.
Detailed metric definitions are in \Apx{sec:app_metrics}.

\paragraph{Implementation.}
We use Qwen2.5-7B-Instruct~\citep{qwen2.5} as the frozen target LLM, following TARS~\citep{tars2025} which also uses the Qwen2.5 family.
The compliance point is defined as the first turn where a majority (at least two of three) of the consensus classifiers flag the response as harmful ($\ell_t \geq 0.67$), matching the threshold used during training (see \Apx{sec:app_metrics} for justification and sensitivity analysis).
All results are mean over 5-fold cross-validation.
Hyperparameter selection and training details are in \Apx{sec:app_config}.

\subsection{Main Results}
\label{sec:main_results}

We evaluate whether \proj achieves earlier detection than existing methods while maintaining competitive detection quality.
\Tbl{tbl:main_compact} presents key metrics across all three benchmarks (full results with all metrics in \Apx{sec:app_full_results}).

\begin{table*}[t]
\centering
\caption{\textbf{Main results across three benchmarks.} We report AUROC, F1, FPR, and Lead (turns of early warning before compliance; higher is better). 
AUC here refers to Area Under the ROC Curve (AUROC).
Full results with all metrics are in \Apx{sec:app_full_results}.}
\label{tbl:main_compact}
\small
\setlength{\tabcolsep}{1.8pt}
\renewcommand{\arraystretch}{1.15}
\begin{tabular}{l cccc |cccc |cccc}
\toprule
& \multicolumn{4}{c}{\textbf{XGuard-Train}} & \multicolumn{4}{c}{\textbf{SafeDialBench}} & \multicolumn{4}{c}{\textbf{SafeMTData}} \\
\cmidrule(lr){2-5} \cmidrule(lr){6-9} \cmidrule(lr){10-13}
\textbf{Method} & \textbf{AUC}$\uparrow$ & \textbf{F1}$\uparrow$ & \textbf{FPR}$\downarrow$ & \textbf{Lead}$\uparrow$ & \textbf{AUC}$\uparrow$ & \textbf{F1}$\uparrow$ & \textbf{FPR}$\downarrow$ & \textbf{Lead}$\uparrow$ & \textbf{AUC}$\uparrow$ & \textbf{F1}$\uparrow$ & \textbf{FPR}$\downarrow$ & \textbf{Lead}$\uparrow$ \\
\midrule
\multicolumn{13}{l}{\textit{Guardrail methods (external, stateless)}} \\
WildGuard      & 0.650 & 0.713 & 0.224 & $-$0.53 & 0.688 & 0.625 & 0.177 & $-$0.89 & 0.672 & 0.577 & 0.194 & $-$0.68 \\
R2-Guard       & 0.688 & 0.648 & 0.188 & $-$0.59 & 0.691 & 0.698 & 0.169 & $-$0.64 & 0.675 & 0.591 & 0.137 & $-$0.41 \\
BingoGuard     & 0.752 & 0.639 & 0.176 & $-$0.45 & 0.721 & 0.680 & 0.132 & $-$0.55 & 0.680 & 0.638 & 0.178 & $-$0.36 \\
% LlamaGuard     & 0.613 & 0.561 & 0.169 & $-$0.55 & 0.653 & 0.571 & 0.188 & $-$0.62 & 0.591 & 0.515 & 0.210 & $-$0.65 \\
GuardReasoner  & 0.726 & 0.597 & 0.147 & $-$0.18 & 0.719 & 0.730 & 0.137 & $-$0.23 & 0.691 & 0.534 & 0.139 & $-$0.39 \\
SelfDefend     & 0.818 & 0.616 & 0.220 & $-$0.07 & 0.734 & 0.721 & 0.115 & $-$0.18 & 0.718 & 0.685 & 0.238 & $-$0.35 \\
\midrule
\multicolumn{13}{l}{\textit{Alignment-based methods (modify LLM weights)}} \\
Circuit Breaker & 0.845 & 0.834 & 0.094 & 0.25 & 0.831 & 0.823 & 0.071 & 0.42 & 0.824 & 0.811 & 0.084 & 0.23 \\
X-Boundary     & 0.877 & 0.862 & 0.070 & 0.33 & 0.858 & 0.839 & 0.094 & 0.37 & 0.836 & 0.820 & 0.091 & 0.28 \\
TARS           & 0.895 & 0.876 & \textbf{0.059} & 0.38 & 0.865 & 0.847 & \textbf{0.057} & 0.55 & 0.847 & 0.832 & \textbf{0.042} & 0.31 \\
\midrule
\multicolumn{13}{l}{\textit{Multi-turn stateful detectors (external)}} \\
\proj & \textbf{0.908} & \textbf{0.891} & 0.079 & \textbf{1.11} & \textbf{0.912} & \textbf{0.901} & 0.066 & \textbf{1.20} & \textbf{0.915} & \textbf{0.904} & 0.080 & \textbf{1.06} \\
\bottomrule
\end{tabular}
\end{table*}

\paragraph{Detection timeliness.}
\proj achieves the largest detection lead across all three benchmarks: 1.11 turns on XGuard-Train, 1.20 on SafeDialBench, and 1.06 on SafeMTData.
This means \proj warns more than one full turn before the LLM complies with the jailbreak.
In contrast, all guardrail methods produce negative Lead values (e.g., WildGuard: $-0.53$ on XGuard-Train), detecting attacks only \emph{after} the LLM has already generated harmful content.
Even the strongest alignment-based method, TARS, achieves only 0.38 Lead on XGuard-Train, detecting less than half a turn before compliance.

\paragraph{Detection quality.}
Despite being a purely external module with no LLM weight modification, \proj achieves the highest AUROC (0.908) on XGuard-Train, outperforming TARS (0.895) which requires full weight access.
\proj maintains an FPR of 7.9\%, substantially lower than guardrail methods (e.g., WildGuard: 22.4\%) and competitive with alignment-based methods (TARS: 5.9\%), which benefit from directly modifying the model's refusal behavior.
This low false positive rate is critical for deployment: a safety monitor that frequently interrupts benign conversations is impractical.

\paragraph{Generalization.}
The advantage is consistent across benchmarks with different attack strategies.
On SafeDialBench (7 distinct attack strategies, bilingual), \proj achieves 0.912 AUROC and 1.20 Lead, outperforming the second-best method TARS (0.865 / 0.55).
On SafeMTData (actor-network attacks), \proj achieves 0.915 AUROC and 1.06 Lead versus TARS's 0.847 / 0.31.

\subsection{Ablation Studies}
\label{sec:ablation}

To understand the contribution of each component, we systematically remove one module at a time and measure the impact on detection quality and timeliness (\Tbl{tbl:ablation}).

\begin{table}[t]
\centering
\caption{\textbf{Ablation study on XGuard-Train.} Each row removes one component. Lead: turns before compliance (higher is better).}
\label{tbl:ablation}
\small
\begin{tabular}{lcccc}
\toprule
\textbf{Variant} & \textbf{F1}$\uparrow$ & \textbf{Recall}$\uparrow$ & \textbf{Lead}$\uparrow$ & \textbf{FPR}$\downarrow$ \\
\midrule
Full \proj & 0.891 & 88.7\% & 1.11 & 7.9\% \\
w/o Imagination & 0.825 & 75.4\% & 0.53 & 10.2\% \\
w/o CUSUM & 0.879 & 93.8\% & 1.49 & 13.8\% \\
w/o Concept Cone & 0.740 & 78.8\% & 0.33 & 15.5\% \\
w/o Transition Model & 0.775 & 83.0\% & 0.30 & 9.8\% \\
Attack-only Imagination & 0.889 & 86.2\% & 1.01 & 25.6\% \\
\bottomrule
\end{tabular}
\end{table}

\begin{itemize}[leftmargin=0.2in]
    \item \textbf{Imagination enables early detection.} Removing imagination (\Sect{sec:imagination}) drops Lead from 1.11 to 0.53, confirming that imagination is the key of proactive early detection.
    \item \textbf{CUSUM enables low false alarm rate.} Removing CUSUM (\Sect{sec:cusum}) increases Lead to 1.49, however, FPR surges to 13.8\%, indicating that CUSUM is essential for suppressing false alarms.
    \item \textbf{Concept cone grounds the latent space.} Removing the concept cone (\Sect{sec:world_model}) causes the most severe overall degradation (F1: 0.891$\to$0.740, FPR: 7.9\%$\to$15.5\%), indicating that the cone projection encodes essential safety-relevant knowledge.
    \item \textbf{Transition model enables dynamics.} Removing the transition model (\Sect{sec:world_model}) drops Lead to 0.30 and F1 to 0.775, as the system loses both imagination capability and multi-turn state integration and reduces to single-turn classification.
    \item \textbf{Contrastive design calibrates imagination.} Attack-only imagination (\Sect{sec:imagination}) reduces Lead from 1.11 to 1.01 and increases FPR from 7.9\% to 25.6\%, confirming that comparing against benign futures is necessary to avoid overestimating vulnerability.
\end{itemize}

\begin{figure}[t]
    \centering
    \includegraphics[width=\linewidth]{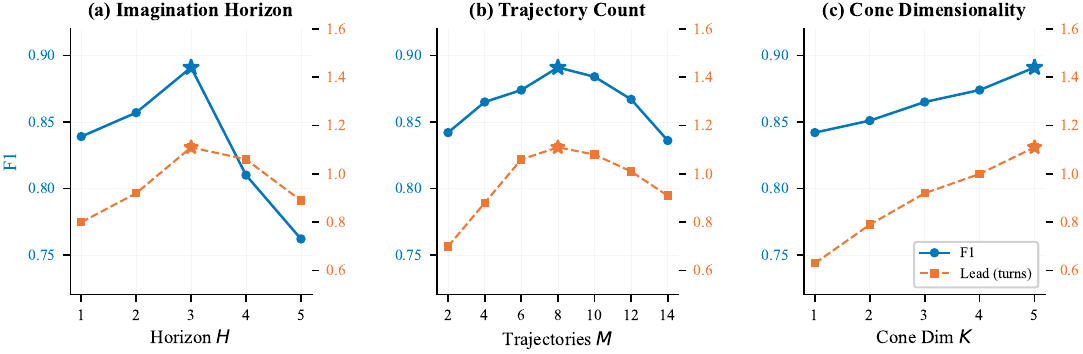}
    \caption{\textbf{Hyperparameter sensitivity.} (a)~Horizon $H{=}3$ maximizes F1 and Lead. (b)~Trajectory count $M{=}8$ saturates the benefit. (c)~Full cone ($K{=}5$) consistently best. Stars mark selected values.}
    \label{fig:hmk_sweep}
\end{figure}

We further sweep the imagination hyperparameters ($H$, $M$) and concept cone dimensionality ($K$).
\Fig{fig:hmk_sweep} shows that $H{=}3$ and $M{=}8$ maximize Lead, with diminishing returns beyond these values, while the full 5-dim cone consistently outperforms lower-dimensional projections.
Additional ablations (transition model architecture) are in \Apx{sec:app_backbone}.

\section{Related Work}
\label{sec:related}
\paragraph{Multi-turn jailbreak attacks and countermeasures.}
Multi-turn jailbreak attacks~\citep{russinovich2025crescendo, fitd2025, star2026} achieve $>$90\% ASR by distributing malicious intent across seemingly innocuous turns.
To counter these threats, guardrail methods~\citep{wildguard2024, r2guard2025, bingoguard2025, guardreasoner2025, selfdefend2025} deploy external classifiers, while alignment-based methods~\citep{circuitbreaker2024, xboundary2025, tars2025} modify LLM weights to internalize safety.
Like these methods, \proj also aims to detect and prevent jailbreaks.
However, guardrails evaluate each turn independently and cannot model the cumulative safety degradation that multi-turn attacks exploit; alignment-based methods require retraining, precluding deployment on closed-source models.
Most critically, both paradigms are \emph{reactive}: they detect harmful content only after it manifests.
\proj addresses all three issues: it operates as an external module (no weight modification), models multi-turn dynamics (not per-turn classification), and detects attacks \emph{before} the LLM complies (proactive rather than reactive).

\paragraph{Safety representations and world models.}
\citet{sva2025} discover a refusal direction in LLM hidden states; RDO~\citep{conceptcone2025} extends this to a multi-dimensional concept cone; SRR~\citep{srr2025} and ReGA~\citep{rega2025} leverage internal representations for safety ranking and runtime safeguarding.
\proj shares with these methods the insight that LLM hidden states encode rich safety-relevant geometric structure.
However, prior work treats these representations as \emph{static} features for single-turn classification or analysis, without modeling how they evolve during a multi-turn conversation.
\proj addresses this gap by learning a \emph{transition model} over safety states, enabling prediction of future safety trajectories.
This dynamics-based approach draws on world models from RL~\citep{dreamerv3, nextlat2025}, which learn environment dynamics and use imagined rollouts to evaluate futures without real interaction.
While world models are typically used for \emph{planning} (selecting optimal actions), \proj adapts them for \emph{detection}: imagining both attack and benign futures to assess whether a conversation is heading toward jailbreak compliance, thus achieves proactive warning with low FPR.
% DeepContext~\citep{deepcontext2026}, a concurrent work, also applies stateful detection to multi-turn dialogues using an RNN.
% However, DeepContext lacks two key capabilities that \proj provides: grounding in verified safety geometry (the concept cone) and contrastive imagination for proactive early warning.

\section{Conclusion}
\label{sec:conclusion}
We proposed \proj, a lightweight external module that requires no LLM weight modification.
\proj formulates the \emph{proactive early jailbreak detection} problem and introduces a new metric, \emph{detection lead}, to measure detection timeliness.
It combines a safety state world model for future state prediction, CUSUM for multi-turn evidence accumulation, and contrastive imagination for proactive early warning.
On three benchmarks against 8 baselines, \proj achieves the best detection timeliness (1.06--1.20 turns before compliance) while maintaining competitive false positive rates, outperforming all baselines in detection quality.
Ablation studies confirm that each component is effective: contrastive imagination drives early detection, CUSUM suppresses false alarms, and the concept cone grounds the safety latent space.
Limitations are discussed in \Apx{sec:app_limitations}.

\bibliography{references}
\bibliographystyle{colm2026_conference}

\appendix
% ============================================================
% Appendix A: Method Details
% ============================================================
\newpage
\section{Method Details}
\label{sec:app_method}

\subsection{Notation Summary}
\label{sec:app_notation}

\Tbl{tbl:notation_full} summarizes all notation used in this paper.

\begin{table}[ht]
\centering
\caption{Full notation summary.}
\label{tbl:notation_full}
\small
\begin{tabular}{cl}
\toprule
\textbf{Symbol} & \textbf{Description} \\
\midrule
$o_t$ & Conversation prefix up to user turn $t$ (before $y_t$) \\
$\bh_t$ & Hidden state from layer 19 of frozen LLM at turn $t$ \\
$\bs_t$ & Concept cone projection ($K$-dim safety signature) \\
$\bz_t$ & Full safety latent state ($[\bs_t \;;\; \text{XAttn}_\phi(\bs_t, \bh_t)]$) \\
$\ba_t$ & Encoded user action at turn $t$ \\
$e_\phi$ & Safety state encoder: $o_t \to \bz_t$ \\
$\text{XAttn}_\phi$ & Cross-attention extension: $(\bs_t, \bh_t) \to \mathbb{R}^{d_{\text{ext}}}$ \\
$g_\theta$ & Safety-State Transformer (transition model) \\
$\psi$ & Action encoder: $u_t \to \ba_t$ \\
$d_\omega$ & Safety discriminator: $\bz_t \to r_t$ \\
$r_t$ & Risk score from discriminator $d_\omega(\bz_t)$ \\
$\mathrm{logit}_\epsilon(\cdot)$ & Clipped log-odds: $\log(\mathrm{clip}(r,\epsilon,1{-}\epsilon)/(1{-}\mathrm{clip}(r,\epsilon,1{-}\epsilon)))$ \\
$\Lambda_t$ & Clipped log-odds of $r_t$, i.e., $\mathrm{logit}_\epsilon(r_t)$ \\
$\Gt$ & CUSUM statistic at turn $t$ \\
$\Vt$ & Vulnerability score from contrastive imagination \\
$A, \kappa, \tau$ & CUSUM threshold, tolerance, imagination trigger ratio \\
$A_{\text{imag}}$ & Vulnerability threshold for proactive warning \\
$M$ & Number of sampled imagination trajectories per condition \\
$H$ & Imagination rollout horizon (number of future steps) \\
$\ell_t^{(n)}$ & Turn-level graduated safety label ($\ell_t \in \{0, \tfrac{1}{3}, \tfrac{2}{3}, 1\}$) \\
$c$ & Compliance point (first harmful LLM response) \\
$N$, $T_n$ & Number of training conversations; turns in conversation $n$ \\
\bottomrule
\end{tabular}
\end{table}

\subsection{Architecture Details}
\label{sec:app_architecture}

\paragraph{Why Transformer over RNN.}
Multi-turn attack strategies involve long-range dependencies: gradual escalation, strategy switching, and topic drift span many turns.
The self-attention mechanism naturally captures dependencies between arbitrary turn pairs.
\citet{nextlat2025} confirm that Transformers can learn compact latent world models effectively.

\paragraph{Layer selection.}
We select layer 19 by running the Refusal Direction Optimization (RDO) algorithm~\citep{conceptcone2025} on harmful and harmless prompts from SALAD-Bench~\citep{saladbench2024}, sweeping over layers and selecting the one that yields the highest refusal manipulation effectiveness for Qwen2.5-7B.
This layer choice is model- and scale-specific and must be re-determined for other architectures.

\paragraph{Cross-attention details.}
The safety signature $\bs_t \in \mathbb{R}^5$ (detached from the computation graph) is linearly projected to produce query vectors $Q = W_Q \bs_t$.
The full hidden state $\bh_t \in \mathbb{R}^{3584}$ is partitioned into a sequence of fixed-size chunks, and each chunk is linearly projected to produce keys $K$ and values $V$.
Standard cross-attention retrieves the most relevant features: $\text{XAttn}_\phi(\bs_t, \bh_t) = \text{softmax}(QK^\top / \sqrt{d_k}) V \in \mathbb{R}^{d_{\text{ext}}}$.

\paragraph{Transition model details.}
The Transformer $g_\theta$ has 4 layers, 4 attention heads, model dimension 128, and Pre-LN normalization.
Since $\bz_t \in \mathbb{R}^{69}$ and $\ba_t \in \mathbb{R}^{64}$ are concatenated, a linear layer projects the 133-dimensional input to the internal dimension of 128; output tokens are projected back to 69 dimensions.

\subsection{Imagination Procedure Details}
\label{sec:app_imagination_procedure}

\paragraph{Action pool construction.}
The attack pool $\mathcal{A}_{\text{atk}}$ and benign pool $\mathcal{A}_{\text{ben}}$ are constructed from the training fold of each cross-validation split.
We encode every user turn from attack conversations as $\ba_t = \psi(u_t)$ and collect them into $\mathcal{A}_{\text{atk}}$; likewise, all user turns from benign conversations form $\mathcal{A}_{\text{ben}}$.
All turns are included regardless of their position in the conversation (i.e., both early benign-looking turns and late escalation turns from attack conversations enter $\mathcal{A}_{\text{atk}}$).
We use uniform random sampling from these pools during imagination.
Importantly, the pools are reconstructed for each cross-validation fold using only training data, ensuring no test-fold leakage.
The pool operates entirely in the 64-dimensional action embedding space, so imagination does not require decoding back to text or invoking the LLM.

\paragraph{Rollout procedure.}
For each condition $\gamma \in \{\text{atk}, \text{ben}\}$, we roll out $M{=}8$ independent trajectories of horizon $H{=}3$, each initialized from the current observed state ($\hat{\bz}_{t}^{(k,\gamma)} = \bz_t$, $\hat{G}_{t}^{(k,\gamma)} = \Gt$).
At each imagined step $j = 1, \ldots, H$, we sample an action $\ba_{t+j}^{(k,\gamma)} \sim \mathcal{A}_\gamma$ uniformly at random, predict the next safety state:
\begin{equation}
    \hat{\bz}_{t+j}^{(k,\gamma)} = g_\theta\!\big(\underbrace{(\bz_0, \ba_1), \ldots, (\bz_{t-1}, \ba_t)}_{\text{real context}},\; \underbrace{(\hat{\bz}_{t}^{(k,\gamma)}, \ba_{t+1}^{(k,\gamma)}), \ldots, (\hat{\bz}_{t+j-1}^{(k,\gamma)}, \ba_{t+j}^{(k,\gamma)})}_{\text{imagined context}}\big),
\end{equation}
score it via $\hat{r}_{t+j}^{(k,\gamma)} = d_\omega(\hat{\bz}_{t+j}^{(k,\gamma)})$, and update the imagined CUSUM:
\begin{equation}
    \hat{G}_{t+j}^{(k,\gamma)} = \max\!\big(0,\; \hat{G}_{t+j-1}^{(k,\gamma)} + \mathrm{logit}_\epsilon(\hat{r}_{t+j}^{(k,\gamma)}) - \kappa\big).
\end{equation}

\subsection{Imagination Loss Details}
\label{sec:app_imagination_loss}

The imagination loss encourages the discriminator to separate attack from benign imagined states:
\begin{equation}
    \mathcal{L}_{\text{imag}} = \underbrace{\max\!\big(0,\; \delta - (r_{\text{atk}} - r_{\text{ben}})\big)}_{\text{margin}} + \lambda_{\text{cal}} \underbrace{\big[\mathrm{BCE}(r_{\text{atk}}, 1) + \mathrm{BCE}(r_{\text{ben}}, 0)\big]}_{\text{calibration}},
\end{equation}
where $r_{\text{atk}} = d_\omega(\mathrm{sg}(\hat{\bz}_{t+H}^{\text{atk}}))$, $r_{\text{ben}} = d_\omega(\mathrm{sg}(\hat{\bz}_{t+H}^{\text{ben}}))$, $\delta = 0.3$, and $\lambda_{\text{cal}} = 0.5$.
The margin term ensures a clear gap between attack and benign risk scores (directly determining $\Vt$), while the calibration term anchors absolute values to prevent degenerate solutions where both scores are high (false alarms) or both low (missed detections).

\subsection{Training Configuration}
\label{sec:app_config}

We use AdamW with learning rate $5 \times 10^{-4}$, weight decay $10^{-4}$, batch size 128, up to 200 epochs with 5-fold cross-validation and early stopping (patience 20 on validation safety loss).
Gradient clipping is set to 1.0.
We use CosineAnnealing learning rate scheduling with $T_{\max} = 200$.
Loss weights: $\lambda_1 = 1.0$ (discrimination), $\lambda_2 = 0.5$ (imagination).
The margin in $\mathcal{L}_{\text{imag}}$ is $\delta = 0.3$ with calibration weight $\lambda_{\text{cal}} = 0.5$.

\paragraph{Gradient routing.}
The cone projections ($\bs_t$) are hardcoded with no gradient.
The transition model $\theta$ receives gradients from $\mathcal{L}_{\text{trans}}$ only; input states $\bz_{t-1}$ are detached so gradients do not flow back into $\phi$.
The cross-attention parameters $\phi$ are updated through $\mathcal{L}_{\text{disc}}$ via $d_\omega(\bz_t)$, since $\bz_t$ depends on $\text{XAttn}_\phi$.
The discriminator $\omega$ receives gradients from both $\mathcal{L}_{\text{disc}}$ and $\mathcal{L}_{\text{imag}}$; imagined states are stop-gradiented, making $\omega$ the only parameter updated by the imagination loss.

% ============================================================
% Appendix B: Experimental Setup
% ============================================================
\section{Full Experimental Setup}
\label{sec:app_setup}

\subsection{Dataset Statistics}
\label{sec:app_dataset}

We use XGuard-Train~\citep{xteaming2025} as our primary evaluation dataset. XGuard-Train contains 30,695 multi-turn jailbreak conversations spanning 13 risk categories, generated via the X-Teaming multi-agent framework with diverse attack strategies. The average dialogue length is around 5 turns.

\paragraph{Topic-matched benign construction.}
For each attack dialogue, we extract the topic from the first user turn and prompt GPT-4o to generate a multi-turn benign conversation on the same topic with a similar number of turns.
To ensure quality, we verify with all three labeling classifiers (HarmBench, GPT-4o, MD-Judge) that no turn in the generated benign conversation is flagged as harmful; any flagged conversation is regenerated.
This topic-matching procedure ensures that the only distinguishing signal between attack and benign conversations is the \emph{trajectory dynamics} (gradual safety erosion), not the surface-level topic.
We apply the same procedure to all three benchmarks (XGuard-Train, SafeDialBench, SafeMTData).

% \begin{table}[ht]
% \centering
% \caption{Dataset statistics for XGuard-Train (primary) and SafeMTData (secondary)}. 
% \label{tbl:dataset}
% \begin{tabular}{lcc}
% \toprule
% \textbf{Statistic} & \textbf{XGuard-Train} & \textbf{SafeMTData} & \textbf{SafeDialBench}  \\
% \midrule
% Attack dialogues & 3000 & \fixme{N}& \fixme{N} \\
% Benign dialogues (topic-matched) & 1809 & \fixme{N} & \fixme{N} \\
% Risk categories / Attack families & 13 & \fixme{N} & \fixme{N}\\
% Mean turns per dialogue & 5.36 & \fixme{N}& \fixme{N} \\
% Mean compliance turn & 4.16 & \fixme{N}& \fixme{N} \\
% \bottomrule
% \end{tabular}
% \end{table}

\subsection{Baseline Descriptions}
\label{sec:app_baselines}

We compare against two categories of baselines:

\textit{Guardrail methods} (5): \textbf{WildGuard}~\citep{wildguard2024}, \textbf{R2-Guard}~\citep{r2guard2025}, \textbf{BingoGuard}~\citep{bingoguard2025}, \textbf{GuardReasoner}~\citep{guardreasoner2025}, and \textbf{SelfDefend}~\citep{selfdefend2025}.

\textit{Alignment-based methods} (3): \textbf{Circuit Breaker}~\citep{circuitbreaker2024}, \textbf{X-Boundary}~\citep{xboundary2025}, and \textbf{TARS}~\citep{tars2025}.

\paragraph{Detection protocol for alignment-based methods.}
Alignment-based methods (Circuit Breaker, X-Boundary, TARS) modify LLM weights to resist attacks rather than producing explicit detection signals.
To evaluate them as detectors under our unified protocol, we define detection as follows: at each turn $t$, we feed the conversation prefix $o_t$ to the alignment-modified LLM and check whether its response constitutes a \emph{refusal} (using the refusal detection task of WildGuard~\citep{wildguard2024}).
If the alignment-modified LLM refuses at turn $t$, we record $t$ as the detection turn $t^{\text{det}}$.
If the model complies (produces non-refusal content), the turn is not flagged.
For computing a continuous risk score (needed for AUROC/AUPRC), we use the refusal probability from the alignment-modified LLM's output logits where available, or the WildGuard refusal confidence score otherwise.
Note that this protocol gives alignment-based methods a natural advantage on FPR: because they directly modify the model's refusal behavior, they rarely refuse benign conversations, yielding inherently low false positive rates.
This explains why TARS achieves the lowest FPR in our evaluation despite being designed for safety alignment rather than detection.

\paragraph{Detection protocol for guardrail methods.}
Each guardrail evaluates the user prompt and/or LLM response at every turn independently.
We use each guardrail's default prompt template and harmfulness task configuration.
A turn is flagged as harmful if the guardrail's output exceeds its default decision threshold.
The detection turn $t^{\text{det}}$ is the \emph{first} flagged turn in the conversation.
For AUROC/AUPRC computation, we use each guardrail's continuous confidence score (probability or logit) as the per-conversation risk score, taking the maximum across all turns.
Since guardrails evaluate each turn independently without modeling multi-turn dynamics, they can only detect attacks \emph{after} harmful content manifests, resulting in negative Lead values.

\paragraph{Guardrail model versions.}
WildGuard: \texttt{allenai/wildguard} (7B, response harmfulness task).
R2-Guard: official implementation with MLN reasoning~\citep{r2guard2025}.
BingoGuard: \texttt{llama3.1-8B} variant, 5-level severity grading (levels $\geq$3 counted as harmful)~\citep{bingoguard2025}.
GuardReasoner: 8B model with chain-of-thought reasoning enabled~\citep{guardreasoner2025}.
SelfDefend: fine-tuned Llama-2-7B with dual-prompt pipeline (P\_direct + P\_intent)~\citep{selfdefend2025}.

\paragraph{Alignment-based model training on Qwen2.5-7B.}
All alignment-based methods were retrained on Qwen2.5-7B-Instruct following each method's official training procedure and publicly released code:
Circuit Breaker uses representation rerouting with the retain set from \citet{circuitbreaker2024} adapted to Qwen's layer structure.
X-Boundary follows the official code release with both single-turn and multi-turn safety training data~\citep{xboundary2025}.
TARS uses the $\lambda{=}0.5$ harmful/harmless reward mixture with GRPO on Qwen-2.5-7B-Instruct~\citep{tars2025}.

\paragraph{Labeling classifier details.}
Turn-level labels are produced by HarmBench~\citep{harmbench2024} (fine-tuned Llama-2-13B classifier), GPT-4o~\citep{gpt4o2024} (\texttt{gpt-4o-2024-08-06}), and MD-Judge~\citep{saladbench2024} (fine-tuned Mistral-7B).
Each classifier receives only the LLM's response text for the current turn (without conversation history) to ensure per-turn independence.

\subsection{Metric Definitions}
\label{sec:app_metrics}

% \paragraph{Detection quality.} F1, AUPRC, Recall, and FPR at each method's best operating point.

% \paragraph{Timeliness} (at a fixed 5\% FPR operating point, calibrated on the validation set):
\begin{itemize}
    \item \textbf{Lead} (Detection Lead Time): $\text{Lead} = \frac{1}{|\mathcal{D}_{\text{atk}}|}\sum_{i} (c_i - t_i^{\text{det}})$. Higher is better; negative values indicate detection after compliance.
    \item \textbf{EWR} (Early Warning Rate): fraction of attack dialogues where \proj detects earlier than the baseline.
\end{itemize}

\paragraph{Compliance point and label design.}
Defining what constitutes a ``successful'' jailbreak---and at which turn it occurs---is a recognized open problem.
\citet{jailbreakeval2024} document that ``there is \emph{surprisingly no consensus} on how to evaluate whether a jailbreak attempt is successful.''
Existing approaches span a wide spectrum:
\emph{conversation-level binary} (Crescendo~\citep{russinovich2025crescendo}: at least one turn flagged by a judge),
\emph{final-turn outcome} (TROJail: reward on last response only),
\emph{per-turn binary} (HarmBench~\citep{harmbench2024}: classify each response independently),
and \emph{multi-turn escalation} (Siege~\citep{siege2025}: tree-search-based multi-turn attacks that exploit partial compliance accumulating across turns).
Siege demonstrates that minor concessions can be systematically exploited to eventually obtain disallowed content.

We adopt a unified labeling framework based on a consensus of three safety classifiers---HarmBench~\citep{harmbench2024}, GPT-4o~\citep{gpt4o2024}, and MD-Judge~\citep{saladbench2024}---each evaluating every response independently.
The graduated label averages their outputs and serves \emph{both} as the training signal for the safety discriminator and as the basis for the compliance point definition:
\begin{itemize}
    \item \textbf{Training}: the graduated labels $\ell_t$ capture progressive safety degradation, providing the discriminator with a smooth signal that reflects escalation dynamics rather than a sharp binary transition.
    \item \textbf{Evaluation}: the compliance point $c$ is defined as the first turn at which $\ell_t \geq 0.67$ (a majority of classifiers flag the response as harmful). Detection Lead is then $\text{Lead} = c - t^{\text{det}}$. Using the same threshold for training and evaluation ensures consistency: the model is trained to detect turns that the evaluation deems harmful.
\end{itemize}
This unified design ensures that training and evaluation operate on the same notion of safety degradation.
\textbf{None of these three classifiers appears as an evaluation baseline} in \Tbl{tbl:main_compact}, ensuring complete separation between label sources and evaluation targets.

All metrics are averaged over 5 folds. \Tbl{tbl:std} reports standard deviations for key metrics on XGuard-Train; all differences between \proj and baselines are statistically significant ($p < 0.01$, paired bootstrap with 10,000 resamples).

\begin{table}[ht]
\centering
\caption{\textbf{Standard deviations across 5 folds on XGuard-Train.} All differences between \proj and baselines are significant at $p < 0.01$ (paired bootstrap, 10K resamples).}
\label{tbl:std}
\small
\begin{tabular}{lcccc}
\toprule
\textbf{Method} & \textbf{AUROC} & \textbf{F1} & \textbf{FPR} & \textbf{Lead} \\
\midrule
WildGuard      & .650$\pm$.012 & .713$\pm$.018 & .224$\pm$.015 & $-$0.53$\pm$0.08 \\
TARS           & .895$\pm$.008 & .876$\pm$.011 & .059$\pm$.007 & 0.38$\pm$0.06 \\
\proj          & .908$\pm$.006 & .891$\pm$.009 & .079$\pm$.008 & 1.11$\pm$0.07 \\
\bottomrule
\end{tabular}
\end{table}

\paragraph{Sensitivity to compliance threshold $\theta$.}
Our primary results use $\theta = 0.67$ (majority vote) to define the compliance point.
To verify that our conclusions are robust to this choice, \Tbl{tbl:theta_sweep} reports Lead for \proj and the strongest baselines across a range of thresholds on XGuard-Train.
A lower $\theta$ (e.g., 0.33) places the compliance point earlier, reducing Lead for all methods; a higher $\theta$ (e.g., 1.0) places it later, inflating Lead.
Crucially, \proj maintains the largest Lead at every threshold, confirming that the advantage is not an artifact of the chosen $\theta$.

\begin{table}[ht]
\centering
\caption{\textbf{Sensitivity to compliance threshold $\theta$ on XGuard-Train.} $\theta$: minimum $\ell_t$ to count as compliance. Lead shifts with $\theta$ because the compliance point moves; the key observation is that \proj maintains the largest Lead at every threshold.}
\label{tbl:theta_sweep}
\small
\begin{tabular}{l ccc}
\toprule
\textbf{Method} & $\theta{=}0.33$ (any) & $\theta{=}0.67$ (majority) & $\theta{=}1.00$ (unanimous) \\
\midrule
WildGuard       & $-$3.17  & $-$0.53 & $-$0.12  \\
GuardReasoner   & $-$2.83 & $-$0.18 & 0.25    \\
Circuit Breaker & $-$2.42  & 0.25     & 0.68   \\
TARS            & $-$2.28 & 0.38    & 0.81   \\
\proj           & $-$1.53  & 1.11    & 1.54   \\
\bottomrule
\end{tabular}
\end{table}

% ============================================================
% Appendix C: Extended Results
% ============================================================
\section{Extended Results}
\label{sec:app_results}

\subsection{Full Benchmark Results}
\label{sec:app_full_results}

\Tbl{tbl:main_xguard}, \Tbl{tbl:main_sdb}, and \Tbl{tbl:main_safemt} present the complete results with all metrics on the three benchmarks.

%%% Table: XGuard-Train (Full) %%%
\begin{table*}[ht]
\centering
\caption{\textbf{Full results on XGuard-Train} (primary benchmark; 30K conversations, 13 risk categories, adaptive multi-agent attacks). $\dagger$EWR measures the fraction of dialogues where \proj detects earlier than each baseline; it is undefined when \proj is the reference method.}
\label{tbl:main_xguard}
\resizebox{\textwidth}{!}{
\begin{tabular}{l cc ccccc cc}
\toprule
& \multicolumn{2}{c}{\textbf{Method Properties}} & \multicolumn{5}{c}{\textbf{Detection Quality}} & \multicolumn{2}{c}{\textbf{Timeliness}} \\
\cmidrule(lr){2-3} \cmidrule(lr){4-8} \cmidrule(lr){9-10}
\textbf{Method} & \textbf{Stateful} & \textbf{External} & \textbf{AUROC}$\uparrow$ & \textbf{AUPRC}$\uparrow$ & \textbf{F1}$\uparrow$ & \textbf{Recall}$\uparrow$ & \textbf{FPR}$\downarrow$ & \textbf{Lead}$\uparrow$ & \textbf{EWR}$\uparrow$ \\
\midrule
\multicolumn{10}{l}{\textit{Guardrail methods (external, stateless)}} \\
WildGuard & & \checkmark & 0.650 & 0.595 & 0.713 & 0.726 & 0.224 & -0.53 & 98.1\% \\
R2-Guard & & \checkmark & 0.688 & 0.671 & 0.648 & 0.523 & 0.188 & -0.59 & 98.9\% \\
BingoGuard & & \checkmark & 0.752 & 0.744 & 0.639 & 0.590 & 0.176 & -0.45 & 97.8\% \\
% LlamaGuard & & \checkmark & 0.613 & 0.582 & 0.561 & 0.489 & 0.169 & -0.55 & 98.4\% \\
GuardReasoner & & \checkmark & 0.726 & 0.678 & 0.597 & 0.619 & 0.147 & -0.18 & 97.4\% \\
SelfDefend & & \checkmark & 0.818 & 0.733 & 0.616 & 0.638 & 0.220 & -0.07 & 97.5\% \\
\midrule
\multicolumn{10}{l}{\textit{Alignment-based methods (modify LLM weights)}} \\
Circuit Breaker & & & 0.845 & 0.861 & 0.834 & 0.845 & 0.094 & 0.25 & 94.7\% \\
X-Boundary & & & 0.877 & 0.871 & 0.862 & 0.879 & 0.070 & 0.33 & 94.3\% \\
TARS & & & 0.895 & 0.905 & 0.876 & 0.854 & 0.059 & 0.38 & 93.9\% \\
\midrule
\multicolumn{10}{l}{\textit{Multi-turn stateful detectors (external)}} \\
\proj & \checkmark & \checkmark & \textbf{0.908} & \textbf{0.946} & \textbf{0.891} & \textbf{0.887} & 0.079 & \textbf{1.11} & --\textsuperscript{$\dagger$} \\
\bottomrule
\end{tabular}
}
\end{table*}

%%% Table: SafeDialBench (Full) %%%
\begin{table*}[ht]
\centering
\caption{\textbf{Full results on SafeDialBench} (4K conversations, 7 attack strategies, bilingual). $\dagger$See \Tbl{tbl:main_xguard} caption for EWR note.}
\label{tbl:main_sdb}
\resizebox{\textwidth}{!}{
\begin{tabular}{l cc ccccc cc}
\toprule
& \multicolumn{2}{c}{\textbf{Method Properties}} & \multicolumn{5}{c}{\textbf{Detection Quality}} & \multicolumn{2}{c}{\textbf{Timeliness}} \\
\cmidrule(lr){2-3} \cmidrule(lr){4-8} \cmidrule(lr){9-10}
\textbf{Method} & \textbf{Stateful} & \textbf{External} & \textbf{AUROC}$\uparrow$ & \textbf{AUPRC}$\uparrow$ & \textbf{F1}$\uparrow$ & \textbf{Recall}$\uparrow$ & \textbf{FPR}$\downarrow$ & \textbf{Lead}$\uparrow$ & \textbf{EWR}$\uparrow$ \\
\midrule
\multicolumn{10}{l}{\textit{Guardrail methods (external, stateless)}} \\
WildGuard & & \checkmark & 0.688 & 0.671 & 0.625 & 0.554 & 0.177 & -0.89 & 99.3\% \\
R2-Guard & & \checkmark & 0.691 & 0.690 & 0.698 & 0.607 & 0.169 & -0.64 & 98.5\%  \\
BingoGuard & & \checkmark & 0.721 & 0.774 & 0.680 & 0.606 & 0.132 & -0.55 & 97.6\%  \\
% LlamaGuard & & \checkmark & 0.653 & 0.642 & 0.571 & 0.539 & 0.188 & -0.62 & 98.8\% \\
GuardReasoner & & \checkmark & 0.719 & 0.738 & 0.730 & 0.731 & 0.137 & -0.23 & 97.1\%  \\
SelfDefend & & \checkmark & 0.734 & 0.756 & 0.721 & 0.682 & 0.115 & -0.18 & 96.8\% \\
\midrule
\multicolumn{10}{l}{\textit{Alignment-based methods (modify LLM weights)}} \\
Circuit Breaker & & & 0.831 & 0.860 & 0.823 & 0.802 & 0.071 & 0.42 & 96.3\% \\
X-Boundary & & & 0.858 & 0.868 & 0.839 & 0.830 & 0.094 & 0.37 & 95.9\% \\
TARS & & & 0.865 & 0.870 & 0.847 & 0.819 & 0.057 & 0.55 & 95.6\% \\
\midrule
\multicolumn{10}{l}{\textit{Multi-turn stateful detectors (external)}} \\
\proj & \checkmark & \checkmark & \textbf{0.912} & \textbf{0.940} & \textbf{0.901} & \textbf{0.918} & 0.066 & \textbf{1.20} & --\textsuperscript{$\dagger$} \\
\bottomrule
\end{tabular}
}
\end{table*}

%%% Table: SafeMTData (Full) %%%
\begin{table*}[ht]
\centering
\caption{\textbf{Full results on SafeMTData}~\citep{actorattack2025} (1.7K conversations, actor-network attacks). $\dagger$See \Tbl{tbl:main_xguard} caption for EWR note.}
\label{tbl:main_safemt}
\resizebox{\textwidth}{!}{
\begin{tabular}{l cc ccccc cc}
\toprule
& \multicolumn{2}{c}{\textbf{Method Properties}} & \multicolumn{5}{c}{\textbf{Detection Quality}} & \multicolumn{2}{c}{\textbf{Timeliness}} \\
\cmidrule(lr){2-3} \cmidrule(lr){4-8} \cmidrule(lr){9-10}
\textbf{Method} & \textbf{Stateful} & \textbf{External} & \textbf{AUROC}$\uparrow$ & \textbf{AUPRC}$\uparrow$ & \textbf{F1}$\uparrow$ & \textbf{Recall}$\uparrow$ & \textbf{FPR}$\downarrow$ & \textbf{Lead}$\uparrow$ & \textbf{EWR}$\uparrow$ \\
\midrule
\multicolumn{10}{l}{\textit{Guardrail methods (external, stateless)}} \\
WildGuard & & \checkmark & 0.672 & 0.610 & 0.577 & 0.546 & 0.194 & -0.68 & 97.2\% \\
R2-Guard & & \checkmark & 0.675 & 0.641 & 0.591 & 0.543 & 0.137 & -0.41 & 98.9\% \\
BingoGuard & & \checkmark & 0.680 & 0.691 & 0.638 & 0.561 & 0.178 & -0.36 & 96.7\% \\
% LlamaGuard & & \checkmark & 0.591 & 0.582 & 0.515 & 0.478 & 0.210 & -0.65 & 97.6\% \\
GuardReasoner & & \checkmark & 0.691 & 0.619 & 0.534 & 0.578 & 0.139 & -0.39 & 98.6\% \\
SelfDefend & & \checkmark & 0.718 & 0.702 & 0.685 & 0.591 & 0.238 & -0.35 & 97.6\% \\
\midrule
\multicolumn{10}{l}{\textit{Alignment-based methods (modify LLM weights)}} \\
Circuit Breaker & & & 0.824 & 0.831 & 0.811 & 0.797 & 0.084 & 0.23 & 95.5\% \\
X-Boundary & & & 0.836 & 0.835 & 0.820 & 0.813 & 0.091 & 0.28 & 94.4\% \\
TARS & & & 0.847 & 0.868 & 0.832 & 0.814 & 0.042 & 0.31 & 93.9\% \\
\midrule
\multicolumn{10}{l}{\textit{Multi-turn stateful detectors (external)}} \\
\proj & \checkmark & \checkmark & \textbf{0.915} & \textbf{0.920} & \textbf{0.904} & \textbf{0.897} & 0.080 & \textbf{1.06} & --\textsuperscript{$\dagger$} \\
\bottomrule
\end{tabular}
}
\end{table*}

% Hyperparameter sensitivity tables (H/M sweep and K sweep) omitted here;
% the same information is presented in Fig.~\ref{fig:hmk_sweep} (main text).
\subsection{Imagination and Cone Hyperparameters}
\label{sec:app_imagination}
\label{sec:app_cone}

The hyperparameter sensitivity for imagination horizon $H$, trajectory count $M$, and concept cone dimensionality $K$ is visualized in \Fig{fig:hmk_sweep} (main text).
$H{=}3$ maximizes both F1 and Lead; $M{=}8$ saturates the benefit; the full 5-dim cone ($K{=}5$) consistently outperforms lower-dimensional projections.
$K{=}1$ reduces to the single refusal direction of Arditi et al.~\citep{sva2025}.

\subsection{Transition Model Architecture}
\label{sec:app_backbone}

We compare three architectures for the transition model on attack-vs-benign trajectory classification (\Tbl{tbl:backbone}) using topic-matched benign contrasts that force models to rely on temporal dynamics rather than topic shortcuts.

\begin{table}[ht]
\centering
\caption{\textbf{Temporal modeling is critical.} Backbone comparison on XGuard-Train. $\Delta$AUROC: drop when turn order is shuffled.}
\label{tbl:backbone}
\begin{tabular}{lccc}
\toprule
\textbf{Model} & \textbf{AUROC} & \textbf{Shuffled} & $\boldsymbol{\Delta}$\textbf{AUROC} \\
\midrule
Transformer & 0.908 & 0.633 & 0.275 \\
GRU & 0.856 & 0.608 & 0.248 \\
MLP & 0.517 & 0.410 & 0.107 \\
\bottomrule
\end{tabular}
\end{table}

The Transformer achieves the highest AUROC (0.908) and shows the largest sensitivity to turn order ($\Delta$AUROC = 0.275): shuffling the turn order causes a 27.5\% drop, confirming that the model relies on temporal structure rather than bag-of-turns features.
The MLP, which processes each turn independently, performs near chance (0.517), demonstrating that per-turn features alone are insufficient for multi-turn attack detection.
This validates our choice of a Transformer-based transition model.

\subsection{Computational Overhead}
\label{sec:app_overhead}

\proj adds minimal latency: the Safety-State Transformer forward pass takes $<$5ms per turn, and contrastive imagination (when triggered) adds $<$100ms.
The module occupies negligible GPU memory alongside the frozen 7B LLM.

% ============================================================
% Appendix D: Visualizations
% ============================================================
% \section{Visualizations}
% \label{sec:app_vis}

% \begin{figure}[ht]
%     \centering
%     \includegraphics[width=\linewidth]{cusum_dynamics.png}
%     \caption{\textbf{CUSUM accumulates risk evidence over turns.} The attack conversation (red) enters the gray zone at turn 3, triggering contrastive imagination, and eventually crosses the alarm threshold. The benign conversation (blue) fluctuates near zero, never entering the gray zone.}
%     \label{fig:cusum}
% \end{figure}

% \begin{figure}[ht]
%     \centering
%     \includegraphics[width=0.85\linewidth]{contrastive_imagination.png}
%     \caption{\textbf{Contrastive imagination exposes vulnerability.} From the current state in the gray zone, we simultaneously imagine attack (red) and benign (green) futures. The vulnerability score $\Vt$ is the difference between their average CUSUM values. A large gap indicates the conversation is vulnerable to exploitation.}
%     \label{fig:imagination}
% \end{figure}

% Old radar placeholder removed — replaced by teaser_trio in introduction

% ============================================================
% Appendix: Limitations
% ============================================================
\section{Limitations}
\label{sec:app_limitations}

\proj requires access to the frozen LLM's intermediate representations (layer 19 hidden states), limiting deployment to settings where hidden states are accessible.
The concept cone geometry is model-family-specific and must be re-extracted for each target LLM; validating the anchoring across diverse model families remains future work.
The transition model learns dynamics from specific attack strategies observed during training; zero-shot transfer to entirely unseen attack families may require retraining.
The contrastive imagination samples actions i.i.d.\ per step from the action pools, while real attacks may exhibit strategic correlations that the i.i.d.\ assumption could underestimate.
Additionally, our evaluation relies on three specific safety classifiers to define ground-truth compliance points; disagreements among classifiers or blind spots shared by all three could introduce systematic labeling bias.
Finally, while we evaluate on three diverse benchmarks, all experiments use a single target LLM (Qwen2.5-7B); extending to other model families (e.g., Llama, Mistral) requires validating and re-extracting the concept cone geometry and retraining the transition model, though the lightweight architecture makes this practical.

\end{document}